\documentclass[12pt]{article}
\usepackage[dvips]{graphicx}
\usepackage{pdproc}
\usepackage{pennames}
\makeatletter
\def\@cite#1{[#1]}
\makeatother  
\newcommand{\mhmax}      {\mbox{$m_{\mathrm{h}}{\mathrm{-max}}$}}

\newcommand{\CPC}[3]  {Comp.\ Phys.\ Comm.\ {\bf #1} (#2) #3}

\newcommand{\etal}{{\it et al.}} 

\begin{document}
\renewcommand{\thefootnote}{\alph{footnote}}
\title{
Investigation of the Discovery Potential for Higgs Bosons 
of the Minimal Supersymmetric Extension of the Standard Model (MSSM) with ATLAS}
\author{MARKUS SCHUMACHER}
\address{
Physikalisches Institut, Universit\"at Bonn \\
Nussallee 12, Bonn, 53115, Germany
%%%%% You may comment out the e-mail address line below.  
\\ {\rm E-mail: Markus.Schumacher@physik.uni-bonn.de}}
\abstract{The discovery potential of the ATLAS experiment at the LHC
for Higgs bosons of the MSSM is discussed. Several CP
conserving and one CP violating benchmark scenario are investigated.
}  
\normalsize\baselineskip=15pt
\section{The Higgs sector of the MSSM and Benchmark Scenarios}
\vspace*{-0.3cm}
The Higgs sector of the MSSM contains two Higgs doublets leading 
to five physical Higgs bosons: three neutral and two charged ones.  
At Born level the phenomenology of the Higgs sector is determined by two 
parameters which can be chosen as $\tan\beta$, the ratio of the
vacuum expectation values of the Higgs doublets, and one physical Higgs boson mass.
The lightest neutral Higgs is bounded by the mass of the Z boson
and the Higgs sector is CP conserving at Born level.
However, the Higgs sector receives large radiative corrections which e.g. 
shift the mass of the lightest Higgs boson up to about 133 GeV~\cite{feynhiggs} 
for a top quark mass of 175 GeV
and a SUSY mass scale of 1 TeV.
The masses and couplings depend on additional SUSY parameters appearing at the
level of radiative corrections: $M_0$, $M_2$, $\mu$, $A$ and 
$M_{gluino}$.\footnote{$M_0$ is the common sfermion mass at the electroweak scale,
$M_2$ is the SU(2) gaugino mass parameter at the electroweak scale
and $M_1$, the U(1) gaugino mass parameter, is derived from $M_2$ using the GUT relation
$M_1=M_2(5\sin^2\!\theta_W/3\cos^2\!\theta_W)$, where $\theta_W$ is
the weak mixing angle, $\mu$ denotes the supersymmetric Higgs
mass parameter, $A$ is the common trilinear Higgs-squark coupling
and $M_{gluino}$ denotes the gluino mass.}
In the CP conserving scenarios (CPC) all parameters are real and the
neutral mass eigenstates are equal to the CP eigenstates: 
two CP even, $h,H$ orderd
by mass, and one CP odd, $A$. Only the CP even bosons couple to weak
gauge bosons.
In the CP violating scenarios (CPV) the complex phases related to $A$ 
and $M_{gluino}$ are additional parameters. Non vanishing phases mix the CP
eigenstates to the mass eigenstates $H_1$, $H_2$ and $H_3$, ordered by mass,
to which no well defined CP property can be assigned.
All neutral mass eigenstates may couple to weak gauge bosons and among
each other in these scenarios.

The interpretation of ATLAS Higgs searches is performed in the 
four CPC benchmark scenarios, \mhmax\ scenario, nomixing scenario, gluophobic
scenario, small $\alpha$ scenario, as suggested in \cite{cpcbenchmark} and in one CPV
benchmark scenario, called CPX scenario, as suggested in \cite{cpxbenchmark}.
The five parameters entering at loop level are fixed in the individual
benchmark scenarios and a scan is performed over $\tan\beta$ between 1 and 50 
and the mass of the CP odd Higgs boson $\mathrm{M_A}$ in the CPC scenarios
(the mass of the charged Higgs boson mass $M_{H^\pm}$ in the CPV scenario)
between 50 and 1000 GeV. The phenomenology and parameter sets are explained
in great detail in~\cite{cpcbenchmark} for the CPC case and in~\cite{cpxbenchmark}
for the CPV case.

The \mhmax\ scenario yields the largest values for $M_h$ up to 133 GeV.
The nomixing scenario is similar to the \mhmax\ scenario, but with vanishing
mixing in the stop sector yielding the smallest values for $M_H$ below 116 GeV.
In the gluophobic scenario  the effective coupling of $h$
to gluons is strongly suppressed for a large area of the  $\tan \beta$ and
${\mathrm M_A}$ plane. Therefore the production cross section for
gluon gluon fusion is strongly suppressed. Values of $M_h$ up to 119 GeV are obtained.
In the small $\alpha$ scenario the branching ratio into $b\bar{b}$ and $\tau^+\tau^-$
is suppressed for large $\tan \beta$ and not too large values of $\mathrm{M_A}$.
Values of $M_h$ up to 123 GeV are obtained.
In the CPX scenario, the parameters are chosen in order to maximize the 
CP violating effects in the Higgs sector.
%%%%%%%%%%%%%%%%%%%%%%%%%%%%%%%%%%%%%%%%%%%%%%%%%%%%%%%%%%%%%%%%%%%%%%%%%%%%%%%%%%%%%%%%%%%%%%
\section{Technicalities}
\vspace*{-0.3cm}
The masses of the Higgs bosons, their coupling strength and branching
ratios for all five benchmark scenarios are calculated with 
FeynHiggs (Version 2.1)~\cite{feynhiggs}.\footnote{For the 
CPX scenario the results will be cross checked with the calculations
from CPSUPERH~\cite{cpsuperh} in the near future.}

Leading order cross sections have been used for all production
processes. For the production of neutral Higgs bosons via gluon gluon fusion
(GGF), weak vector boson fusion (VBF), associated production with weak
gauge bosons ($W{\cal H}$)\footnote{$\cal{H}$ here and in the following
denotes a general neutral Higgs mass eigenstate. Only its CP even
component couples to W and Z boson.} 
and heavy quarks ($tt{\cal H}$ and $bb{\cal H}$)
the SM like cross sections have been calculated using the programs 
from reference~\cite{spira} and then applying the appropriate correction
factors to obtain the MSSM cross section values.
The production cross section of charged Higgs bosons via $\mathrm{gb\rightarrow tH^\pm}$ 
has been calculated following~\cite{plehn}. The cross section for
$\mathrm{tt\rightarrow bb W H^\pm}$ has been obtained with PYTHIA~\cite{pythia}.
A top quark mass of 175~GeV\footnote{The mass of the $h$ 
depends linearly on $m_t$, whichs current central value 
is 180 GeV.  Generally a heavier Higgs boson is easier to discover at the LHC.
The discovery potential of the individual channels might change slightly,
but the general conclusions should be robust against an increase of $m_t$.} has been used in all calculations.

The signal efficiencies and number of expected background events
are taken from published ATLAS fast MC studies\footnote{
The following channels have been considered: for neutral Higgs bosons:
VBF with ${\cal H} \rightarrow \tau\tau, WW$ and $\gamma\gamma$~\cite{vbf},
$tt{\cal H}$ with ${\cal H}\rightarrow bb$~\cite{tthbb}, 
${\cal H}\rightarrow \mu\mu$~\cite{bbhmm}
and ${\cal H} \rightarrow \tau\tau$~\cite{bbhta} from GGF and $bb{\cal H}$,
${\cal H}\rightarrow \gamma\gamma$ from GGF, $W,{\cal H}$ and 
$tt{\cal H}$, ${\cal H}\rightarrow ZZ \rightarrow 4 \ell$
and ${\cal H} \rightarrow WW \rightarrow \ell \nu \ell \nu$ from GGF, 
$W{\cal H}$ with ${\cal H} \rightarrow bb$ and ${\cal H}
\rightarrow WW\rightarrow \ell nu \ell nu$, 
$H/A$ with $H/A \rightarrow tt$, $H \rightarrow hh \rightarrow \gamma \gamma bb$ and 
$A \rightarrow Zh \rightarrow \ell \ell bb$ all from~\cite{tdr}.
For the discovery of charged Higgs bosons:
$gb \rightarrow  t H^\pm$ with $H \rightarrow \tau \nu$~\cite{chargedheavy} 
and in the decay of top quarks $pp \rightarrow tt$ with 
$t \rightarrow b H\pm$ and $H \rightarrow tb, \tau \nu$~\cite{chargedlight}.
The VBF channels, ${\cal H} \rightarrow \tau\tau$ decays and charged
Higgs bosons production from t quark decays have only been studied for
luminosity running of the LHC and results are only shown for an
integrated luminosity of 30\,fb$^{-1}$.}.
The key performance figures for e.g. lepton identification and isolation,
b-tagging, $\tau$ identification, trigger efficiencies and mass resolutions have been
obtained from studies using a full simulation of the ATLAS detector.
The effect of almost mass degeneracy of Higgs bosons leading to a
signal overlap and the effect of a larger total decay width of a
Higgs boson when compared to its SM value have been taken into account
when determining the number of expected signal events.
 
Discovery here means that the probability of a background fluctuation
to the number of expected signal+background events
is less than 2.85$\times$10$^{-7}$ using Poissionian statistics.

More details of the updated interpretation can be found in~\cite{ms}.
The LEP exclusion contours shown are taken from~\cite{lephiggs} and~\cite{opalhiggs}.

\section{Discovery Potential in the CPC benchmark scenarios}
\begin{figure}[h!t!b!]
\begin{center}
\includegraphics*[width=5.5cm]{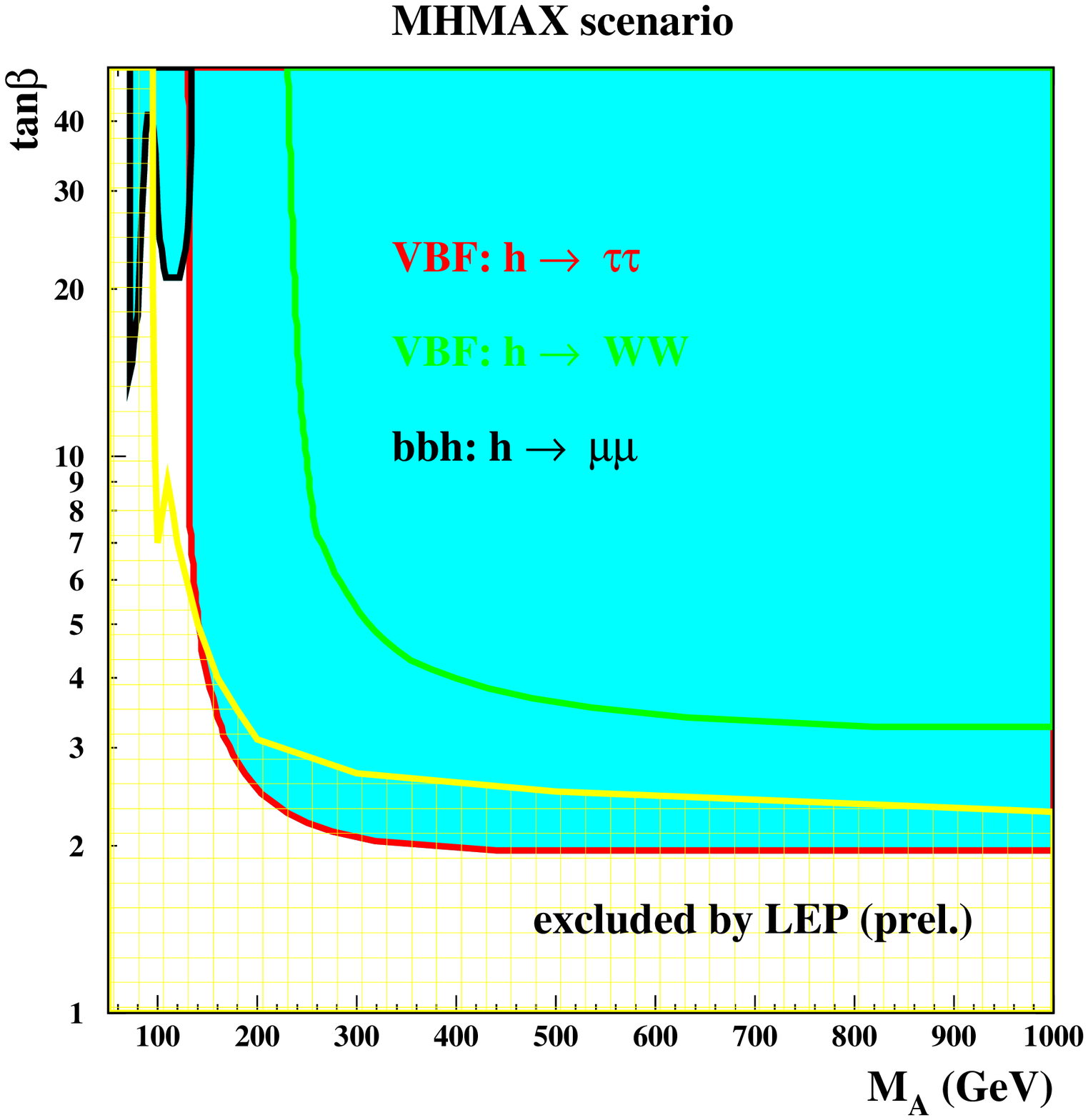}
\includegraphics*[width=5.5cm]{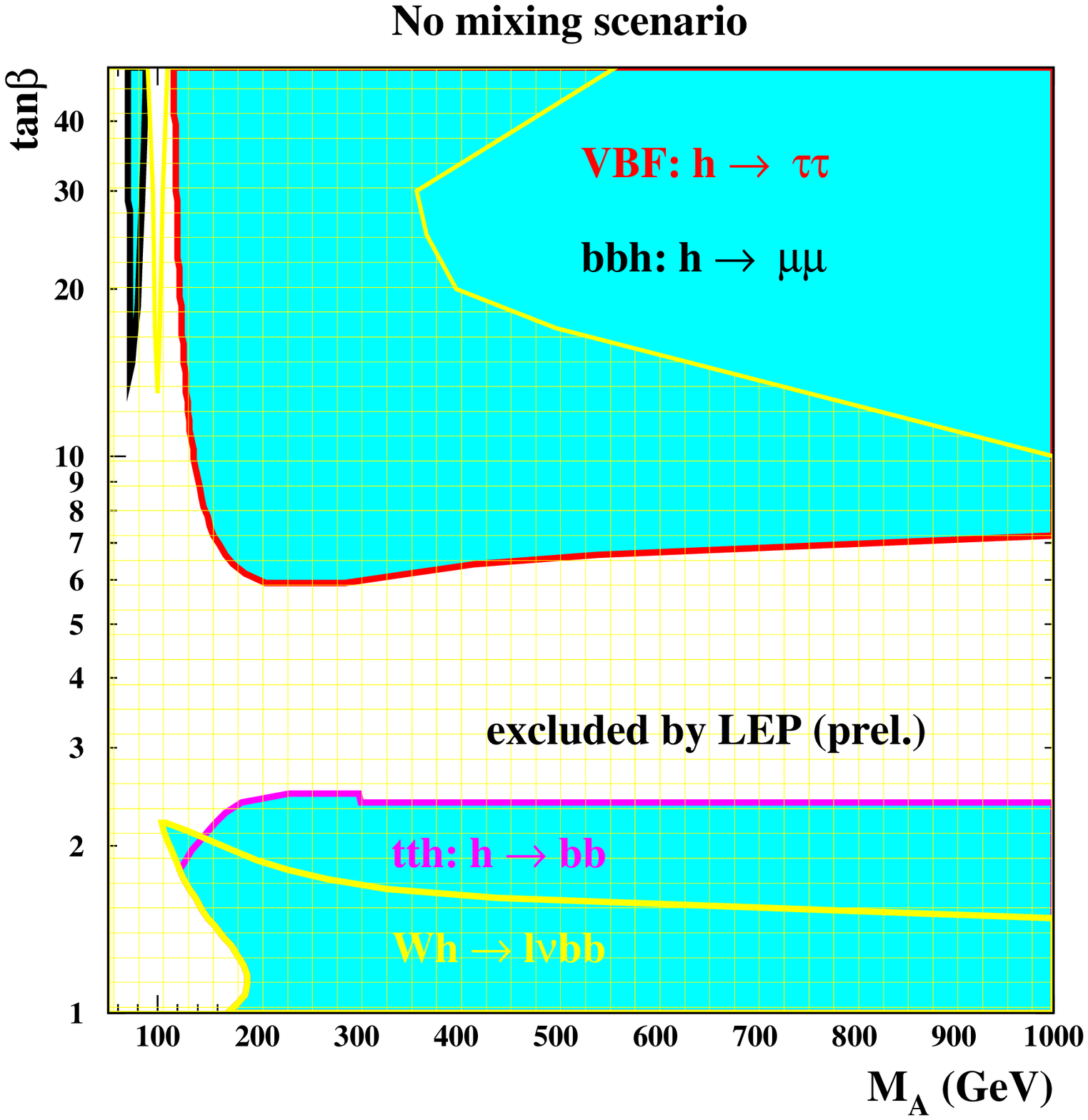}
\includegraphics*[width=5.5cm]{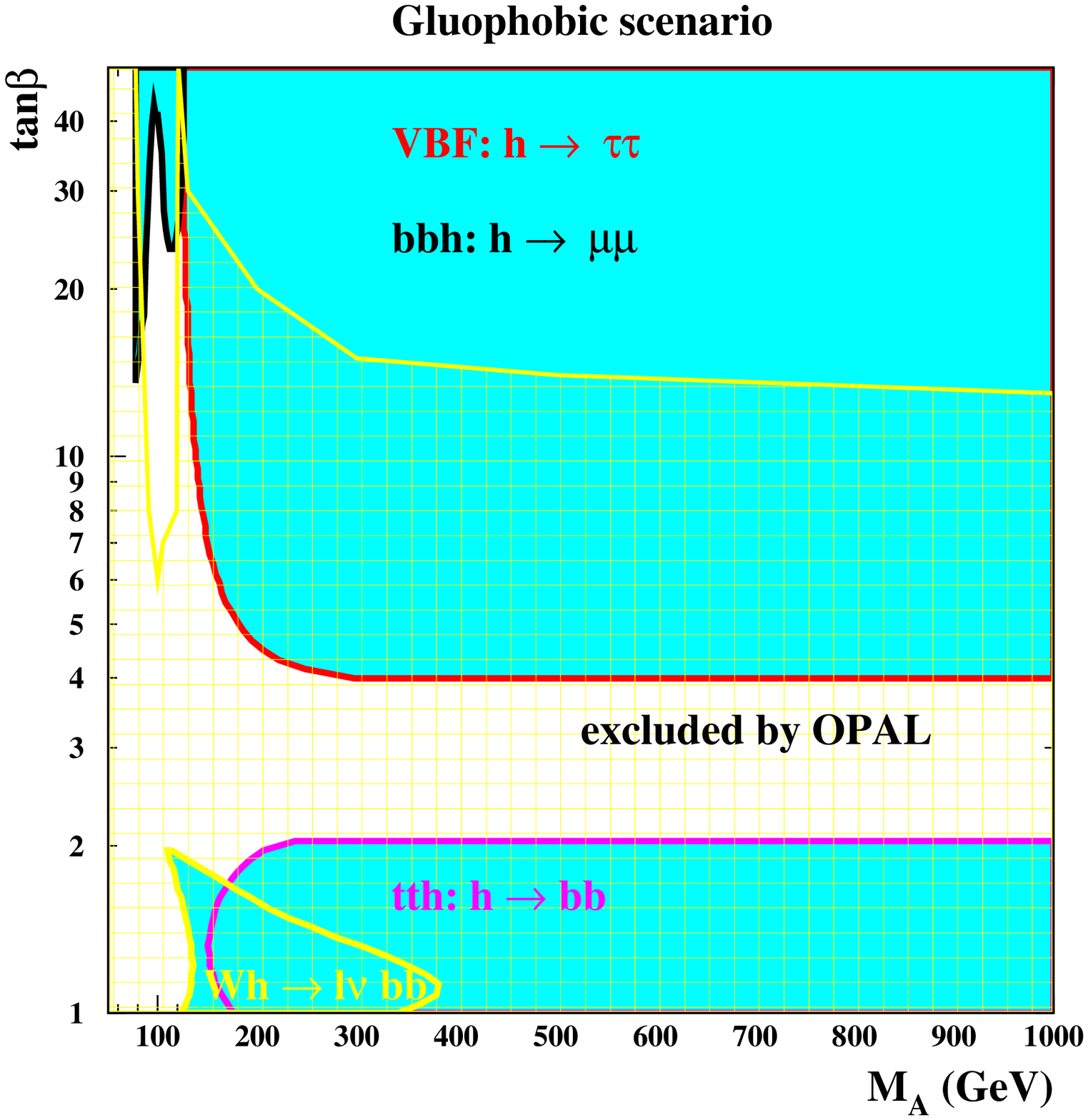}
\includegraphics*[width=5.5cm]{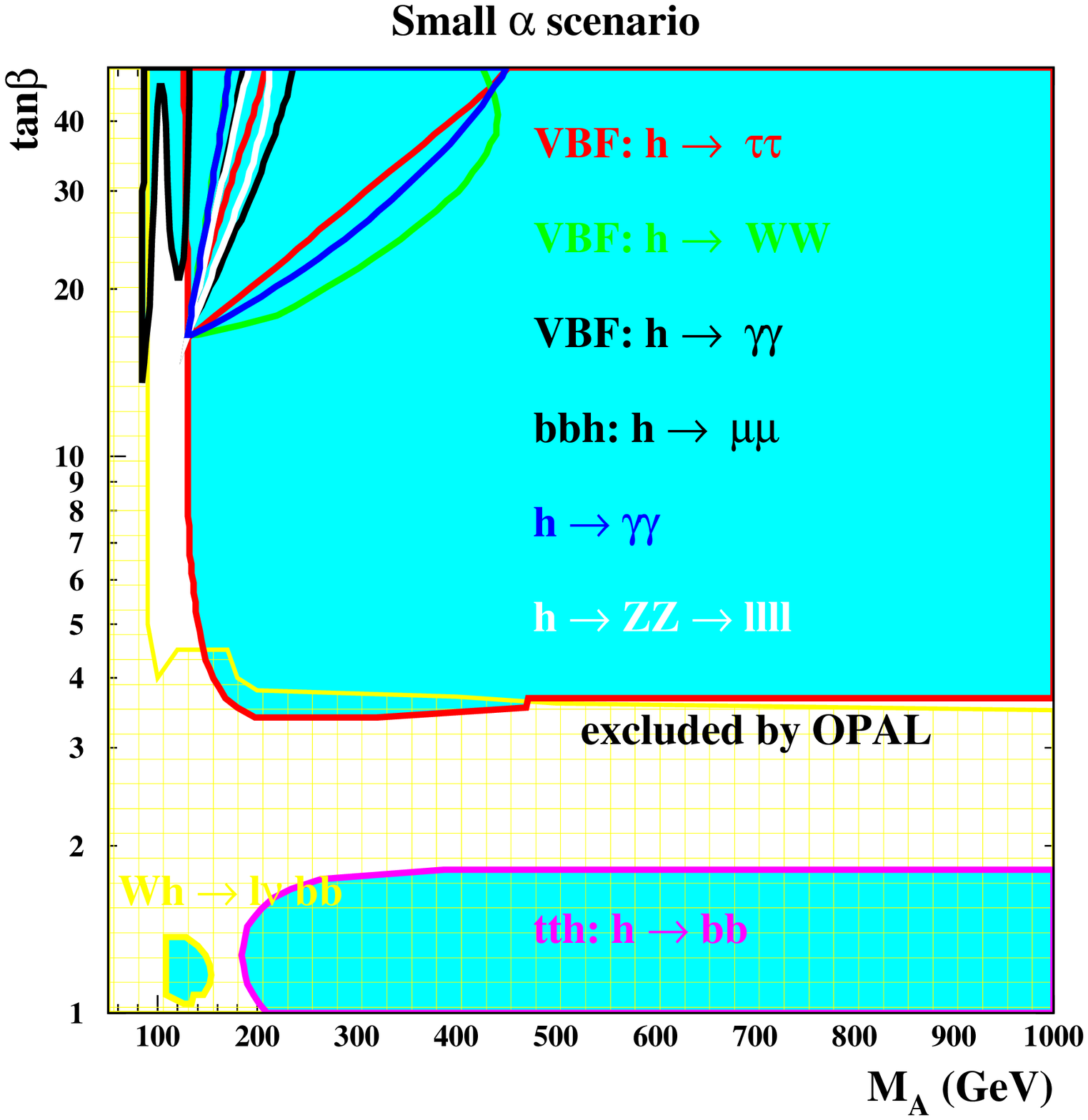}
\caption{Discovery potential for the light CP-even Higgs boson in 
the CPC benchmark scenarios after collecting 30 fb$^{-1}$. 
The cross hatched area is excluded by LEP at 95\% CL.}
\label{fig1}
\end{center}
\end{figure}

\begin{figure}[h!t!b!]
\begin{center}
\includegraphics*[width=5.5cm]{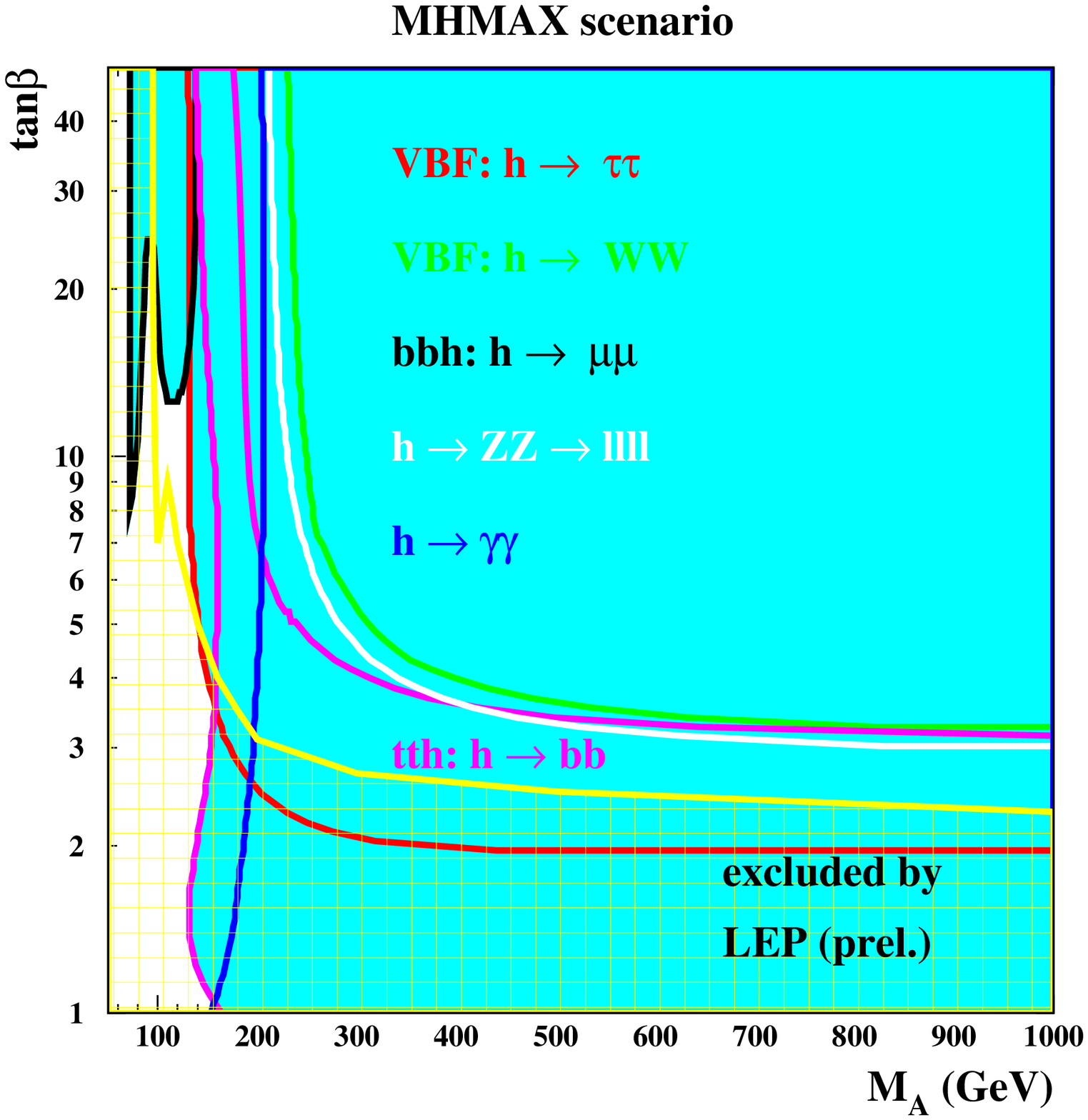}
\includegraphics*[width=5.5cm]{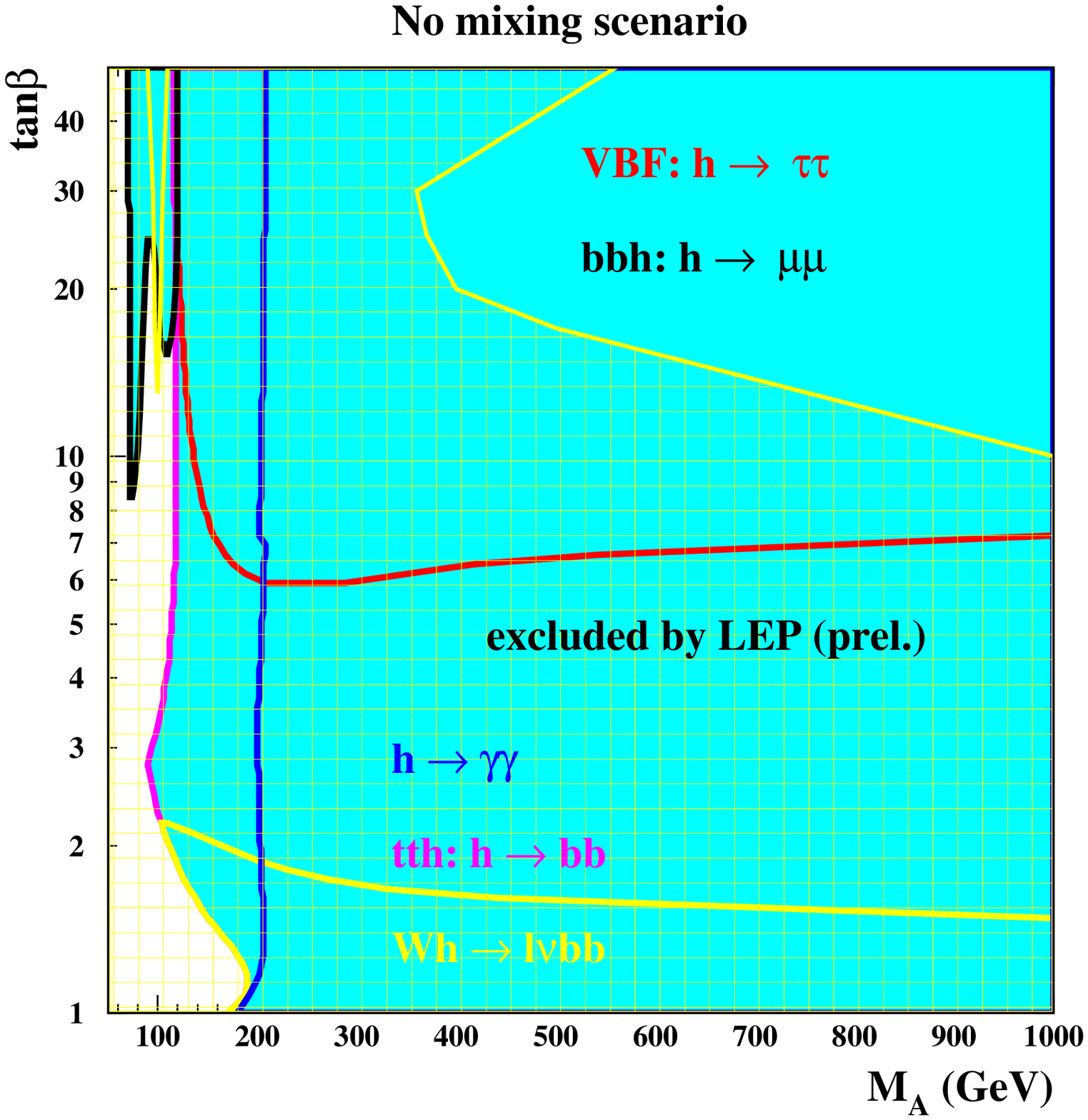}
\includegraphics*[width=5.5cm]{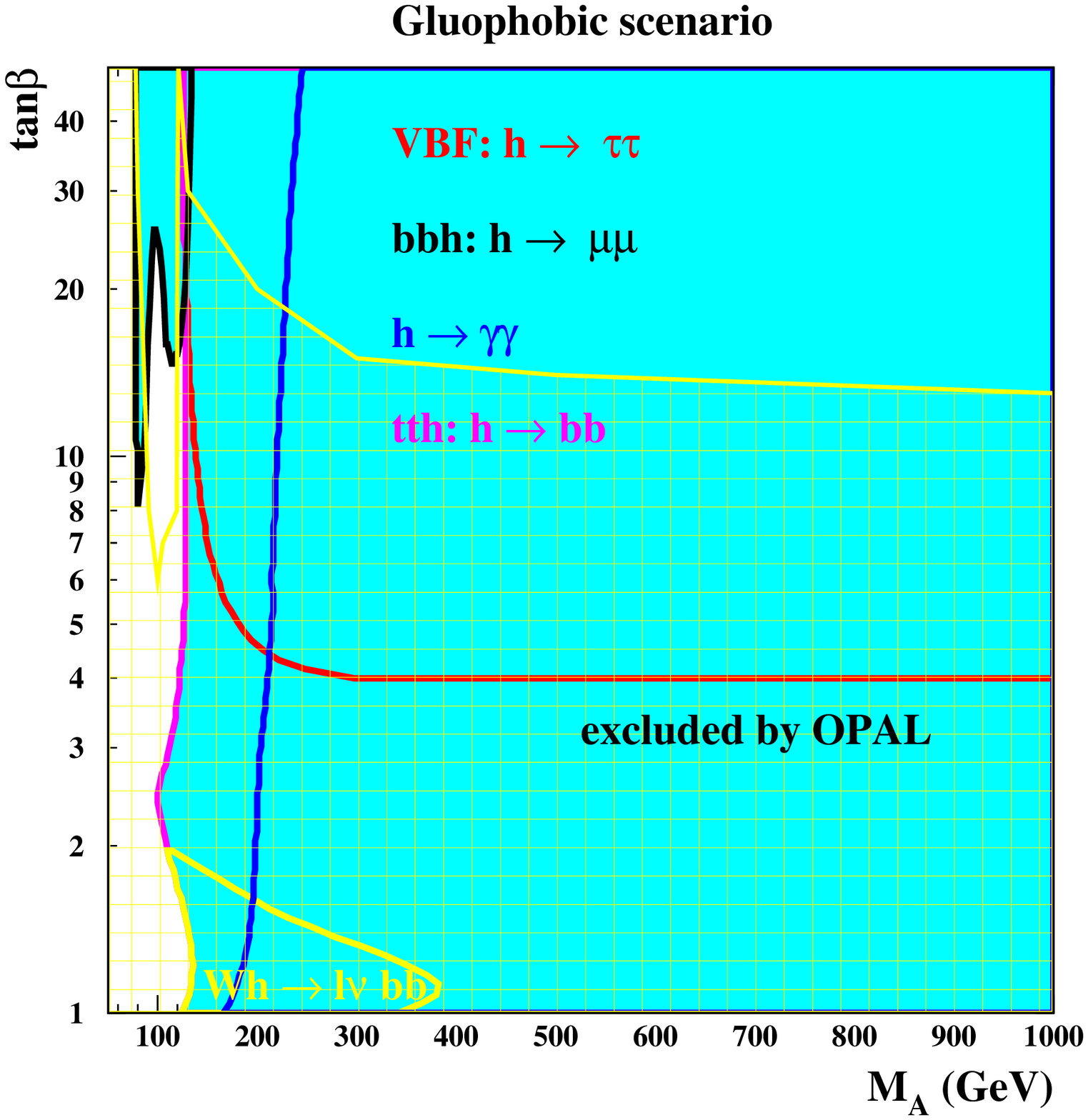}
\includegraphics*[width=5.5cm]{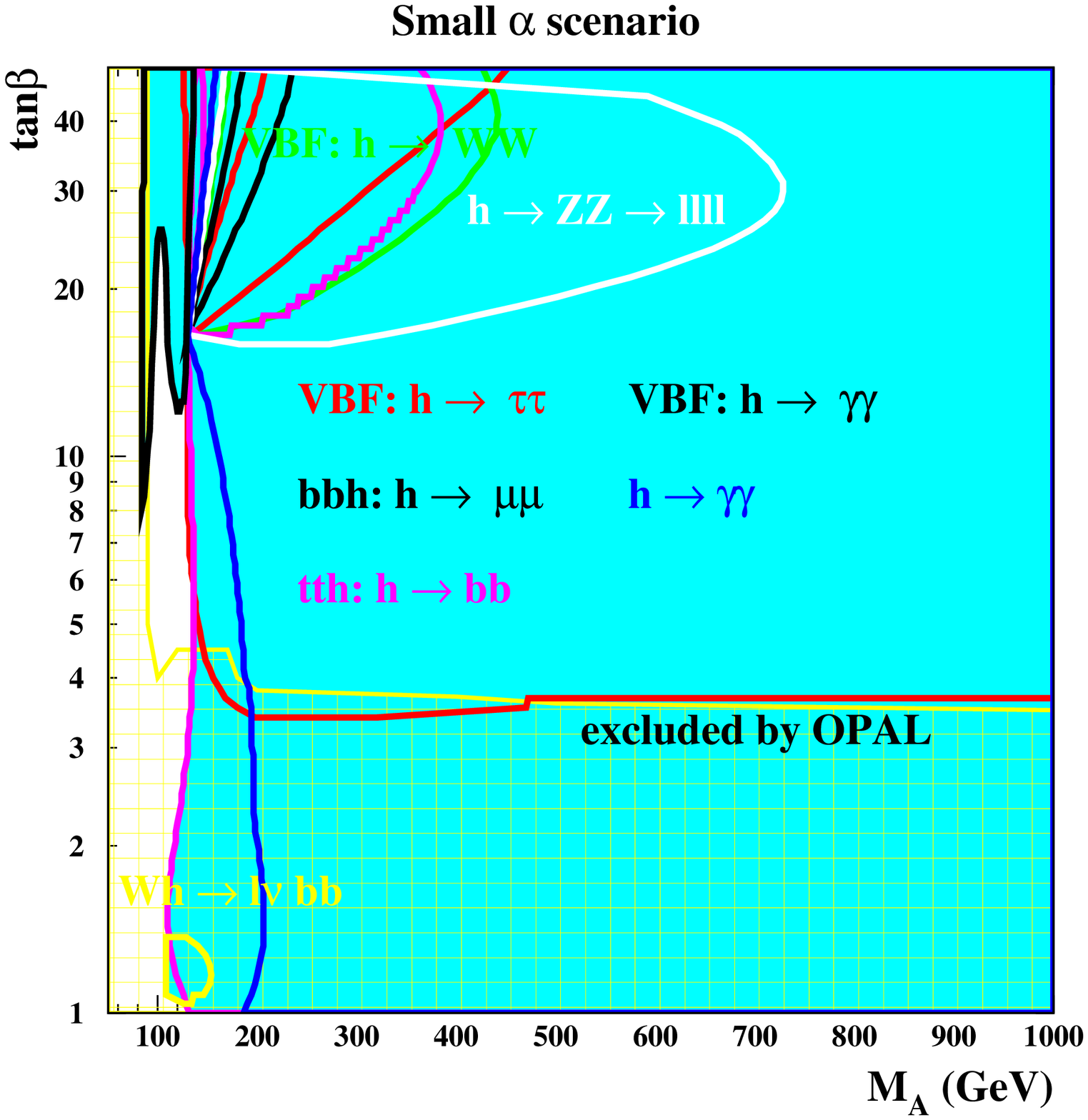}
\caption{Discovery potenital for the light CP-even Higgs boson in 
the CPC benchmark scenarios after collecting 300 fb$^{-1}$.
The cross hatched area is excluded by LEP at 95\% CL.}
\label{fig2}
\end{center}
\end{figure}

\begin{figure}[h!t!b!]
\begin{center}
\includegraphics*[width=5.5cm]{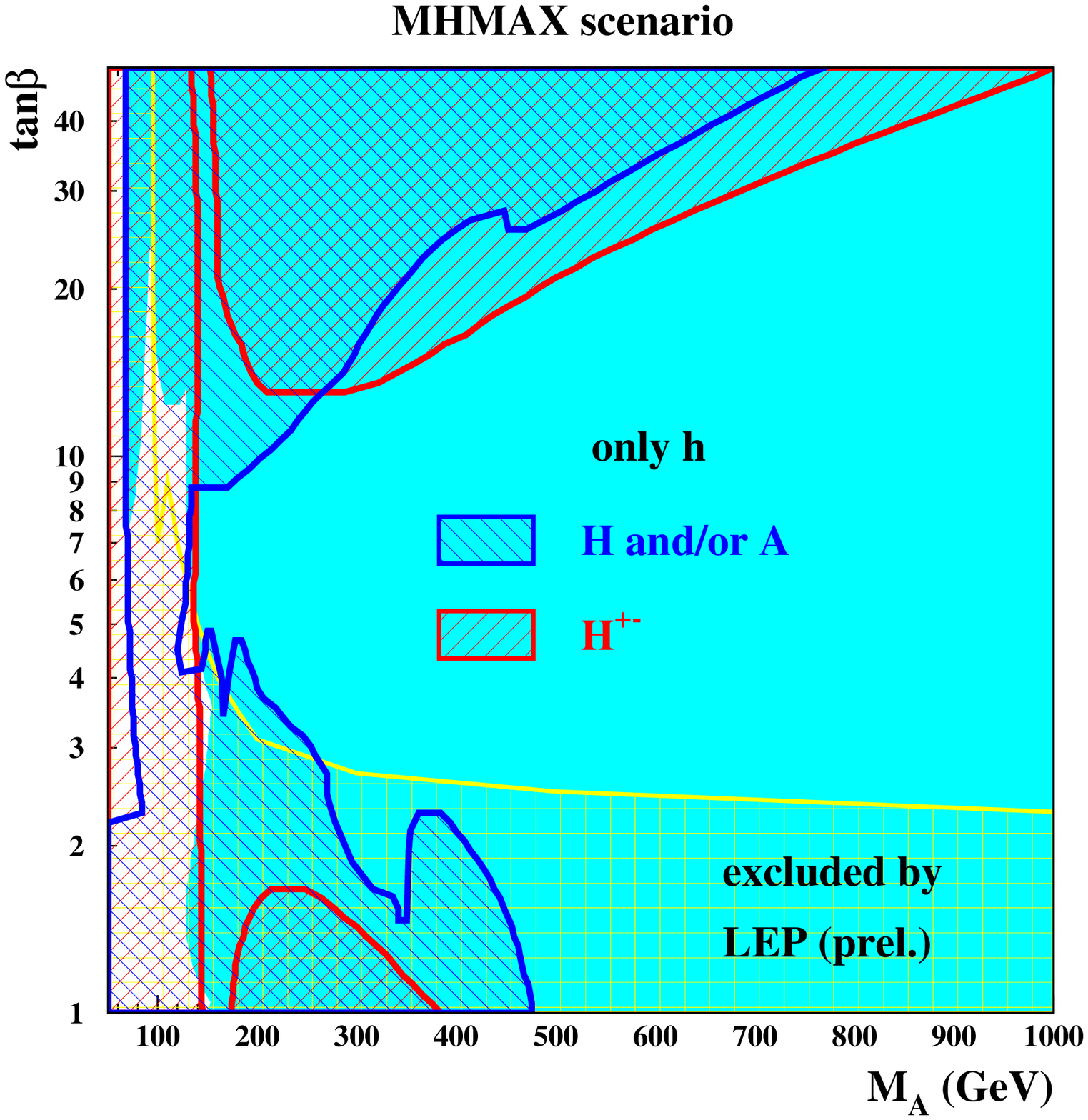}
\includegraphics*[width=5.5cm]{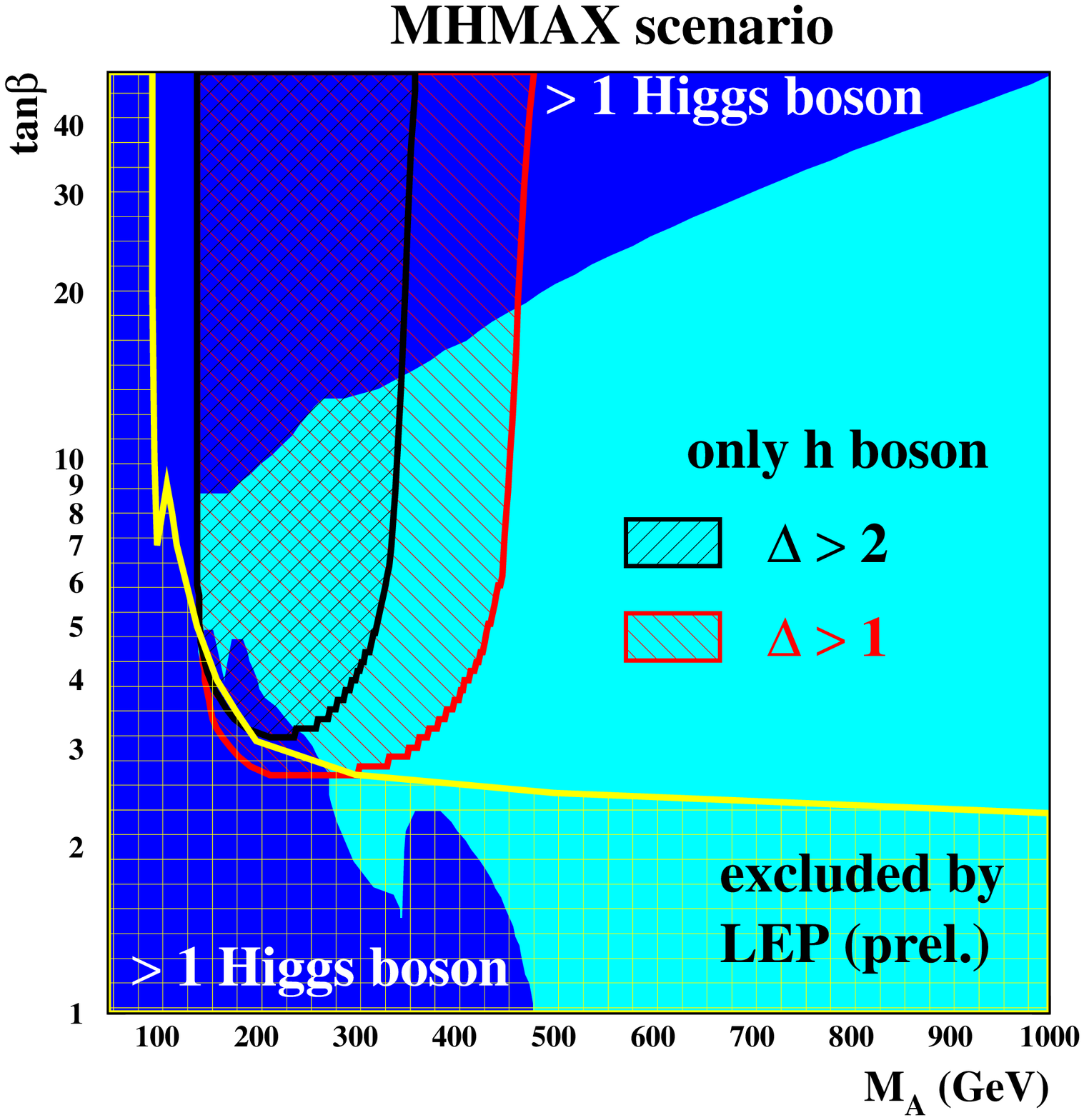}
\caption{Left: overall discovery potential for Higgs bosons in the 
\mhmax\ scenario after collecting 300 fb$^{-1}$. 
The cross hatched area is excluded by LEP at 95\% CL.
Right: sensitivty for
discrimination between SM and MSSM from the measurement of $\Delta$
(see text) in the \mhmax\ scenario.}
\label{fig3}
\end{center}
\end{figure}

\vspace*{-0.3cm}
The discovery potential for the light CP even Higgs boson $h$ 
in the four CPC benchmark scenarios after collecting an integrated
luminosity of 30 (300) fb$^{-1}$ are shown in figure~\ref{fig1}~(\ref{fig2}).
The VBF channel with $h \rightarrow \tau \tau$
dominates the discovery potential at low luminosity running
and covers most of the parameter space left over from the LEP exclusions.
The differences between the \mhmax, no mixing and gluophobic scenario
is mainly due to the fact that in the same point of the parameter
space the mass of the $h$ is different, giving rise to increased or 
decreased sensitivity of the channels under consideration. 
In the small $\alpha$ scenario the effect of surpressed
branching ratios into $\tau$ leptons is visible for $\tan\beta > 20$
and approximately 200 GeV $<\mathrm{M_A}<$ 300 GeV. This hole in the
discovery region 
is nicely complemented by $h$ decays to gauge bosons from VBF or GGF.
For high luminosity running also the channels $h\rightarrow \gamma \gamma$ and
$h \rightarrow ZZ \rightarrow $ 4 leptons and $tth$ with $h\rightarrow bb$
contribute significantly. For all benchmark scenarios in a large part of the
MSSM parameter space discovery is possible via several channels, which  
should allow a determination of parameters of the Higgs sector.
The difference between the phenomenology of the heavy Higgs boson
sector is small especially for larger values of $M_A$, hence only the 
\mhmax\ scenario is discussed below. The discovery potential for the 
heavy neutral Higgs bosons $H$ and $A$ in the not yet by LEP excluded area is given by associated production
with b quarks and the decay into a pair of myons and tau leptons.
Charged Higgs bosons can be observed from top quark decays
for $M_{H^\pm} <$ 170 GeV and from gluon bottom fusion for $M_{H^\pm}
>$ 180 GeV. The overall discovery potential after collecting 300\,fb$^{-1}$ is
shown in figure \ref{fig3} (left). In the whole model parameter space
at least one Higgs boson can be discovered and 
for a significant part of parameter space more than one Higgs boson
can be observed allowing to distinguish between the Higgs sector of
the SM and its MSSM extension via direct observation. However 
a large area at intermediate $\tan\beta$ is left where only the light
Higgs boson $h$ can be discovered. The observation of the light Higgs 
boson $h$ in various search channels might allow a discrimination 
via the measurement of e.g. ratio of branching ratios in the 
same production mode, which is advantageous as several systematic 
uncertainties are cancelled. A first estimate for the sensitivity of
such a discrimination 
between SM and MSSM has been performed using the ratio $R$ of 
the branching ratios measured in the VBF production mode: 
$R = BR(h\rightarrow  \tau\tau)/BR(h\rightarrow WW)$.
The red (black) area in figure\ref{fig3} (right)
indicates this sensitivity
for which $\Delta$, defined as $\Delta = (R_{MSSM}-R_{SM})/\sigma_{exp.}$ 
is larger than 1 (2). Here $\sigma_{exp.}$ denotes the expected error on
the ratio R in this particular point of MSSM parameter space. 
Only statistical uncertainties have been taken into account and it has
been assumed that $M_h$ is measured with high precision. 
Similar results have been obtained in~\cite{duehrssen}.
\vspace*{-0.5cm}

\vspace*{-0.5cm}
\section{Discovery Potenital in the CPX benchmark scenario}
\begin{figure}[h!tb]
\begin{center}
\includegraphics*[width=6cm]{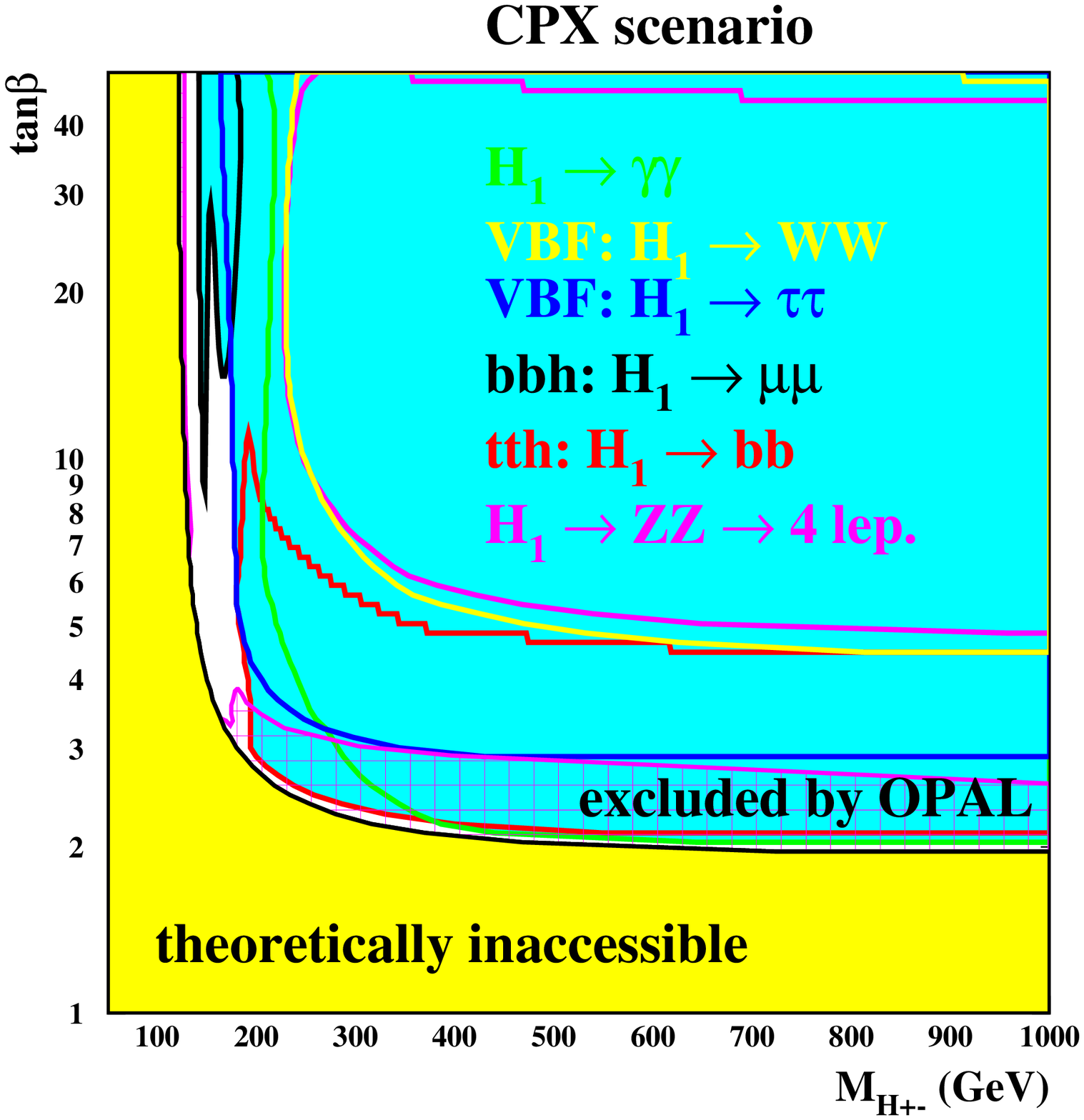}
\includegraphics*[width=6cm]{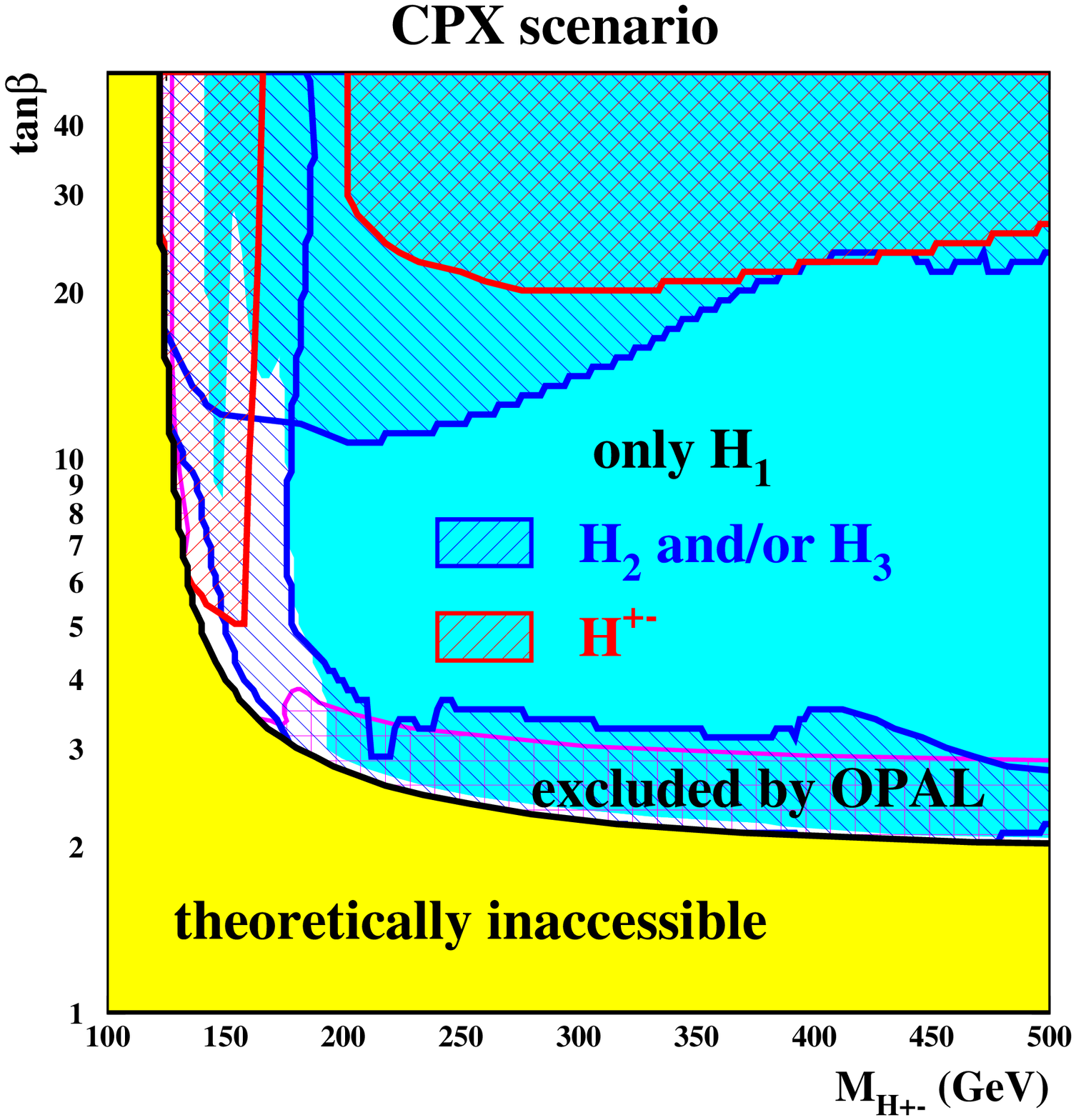}
\caption{Left: discovery potential for the lightest Higgs boson in 
the CPX scenario after collecting 300 fb$^{-1}$. 
The cross hatched area is excluded by LEP at 95\% CL.
Right: overall discovery potential for Higgs bosons in the CPX scenario after 
collecting 300 fb$^{-1}$.}
\label{fig4}
\end{center}
\end{figure}

\vspace*{-0.3cm}
The discovery potential for the lightest neutral Higgs boson $H_1$ 
in the CPX scenario after collecting an integrated
luminosity of 300\,fb$^{-1}$ is shown in figure~\ref{fig4} (left).
The coverage is similar to the CPC scenarios, but the LEP exclusion
is weaker and even low Higgs boson masses $M_{H_1}$ between 0 and 60 GeV
are not excluded yet~\cite{opalhiggs}.
The overall discovery potential for Higgs bosons in the CPX scenario 
after collecting an integrated luminosity of 300\,fb$^{-1}$ is shown 
in figure~\ref{fig4} (right). A region at low ${M_H^\pm}$ and small
$\tan\beta$ remains where no discovery is possible with the channels
and mass ranges investigated so far within the ATLAS collaboration.
In this area the mass of $H_1$ ($H_2$, $H_3$) is in the range 
$<$ 70 GeV (105 to 120 GeV, 140 to 180 GeV).
\vspace*{-0.3cm}
\section{Conclusions}
\vspace*{-0.3cm}
An updated evaluation of the discovery potential of ATLAS Higgs 
searches based on most recent calculations for masses and branching 
ratios in four CPC benchmark scenarios and the CP violating CPX scenario 
has been discussed.
A good discovery potential for the whole MSSM paramter space in the four
CPC benchmark scenarios is expected. The light Higgs boson can be
discovered in multiple search channels, allowing maybe an indirect 
discrimination whether the 
SM or MSSM is realised in nature.
In the CPX scenario an area at low ${M_H^\pm}$ and small
$\tan\beta$ remains which is not covered by today's ATLAS MC studies
which are limited to Higgs boson masses above 70 GeV. Further MC studies are
needed to show whether this area can be covered at the LHC.
\section{Acknowledgements}
\vspace*{-0.3cm}
\small{
I would like to thank S. Heinemeyer,  T. Plehn and M. Spira for providing their
programs and numerous fruitful discussion. This work was partially
supported by the Federal Ministry of Education, Science, Research 
and Technology (BMBF) under contract number 05HA4PD1/5 and
also within the framework of the
German-Israeli Project Cooperation in Future-Oriented Topics (DIP).}
\bibliographystyle{plain}

\end{document}